\pdfoutput=1

\documentclass[11pt]{article}

\usepackage[final]{acl}

\usepackage{times}
\usepackage{latexsym}
\usepackage{listings}
\usepackage[T1]{fontenc}

\usepackage[utf8]{inputenc}

\usepackage{microtype}

\usepackage{inconsolata}

\usepackage{amsmath,amsfonts,bm}

\def\eqref#1{equation~\ref{#1}}

\def\1{\bm{1}}

\DeclareMathAlphabet{\mathsfit}{\encodingdefault}{\sfdefault}{m}{sl}
\SetMathAlphabet{\mathsfit}{bold}{\encodingdefault}{\sfdefault}{bx}{n}

\usepackage{hyperref}
\usepackage{url}
\usepackage{graphicx}
\usepackage{float}
\usepackage{enumitem}

\usepackage{booktabs}
\usepackage{wrapfig}
\usepackage{subcaption}
\usepackage{microtype}

\usepackage{xcolor}
\usepackage{colortbl}
\usepackage{multirow}
\usepackage{tcolorbox}
\usepackage{soul}

\hypersetup{
    colorlinks,
    linkcolor={blue},%
    citecolor={green!60!black},
    urlcolor={blue}%
}

\title{Protecting Users From Themselves: Safeguarding Contextual Privacy in Interactions with Conversational Agents}

\author{Ivoline C. Ngong\thanks{Graduate student at University of Vermont. Work done during summer internship at IBM Research.},~Swanand Kadhe, \textbf{Hao Wang}\thanks{Now at the RedHat AI Innovation Team.}, Keerthiram Murugesan, Justin D. Weisz,\\ 
\textbf{Amit Dhurandhar, Karthikeyan Natesan Ramamurthy} \\
IBM Research. \\
\texttt{kngongiv@uvm.edu,}\\
\texttt{\{swanand.kadhe,keerthiram.murugesan\}@ibm.com,}\\
\texttt{hao-wang@redhat.com},
\texttt{\{jweisz,adhuran,knatesa\}@us.ibm.com}
}

\begin{document}
\maketitle
\begin{abstract}
Conversational agents are increasingly woven into individuals' personal lives, yet users often underestimate the privacy risks associated with them. The moment users share information with these agents —such as large language models (LLMs)— their private information becomes vulnerable to exposure. In this paper, we characterize the notion of contextual privacy for user interactions with LLM-based Conversational Agents (LCAs). It aims to minimize privacy risks by ensuring that users (sender) disclose only information that is both relevant and necessary for achieving their intended goals when interacting with LCAs (untrusted receivers). Through a formative design user study, we observe how even ``privacy-conscious'' users inadvertently reveal sensitive information through indirect disclosures. 
Based on insights from this study, we propose a locally deployable framework that operates between users and LCAs, identifying and reformulating out-of-context information in user prompts. Our evaluation using examples from ShareGPT shows that lightweight models can effectively implement this framework, achieving strong gains in contextual privacy while preserving the user's intended interaction goals. Notably, about 76\% of participants in our human evaluation preferred the reformulated prompts over the original ones, validating the usability and effectiveness of contextual privacy in our proposed framework. We open source the code at  \url{https://github.com/IBM/contextual-privacy-LLM}.
\end{abstract}

\section{Introduction} %

\begin{figure}[t!]
\centering
    \includegraphics[width=1.0\linewidth]{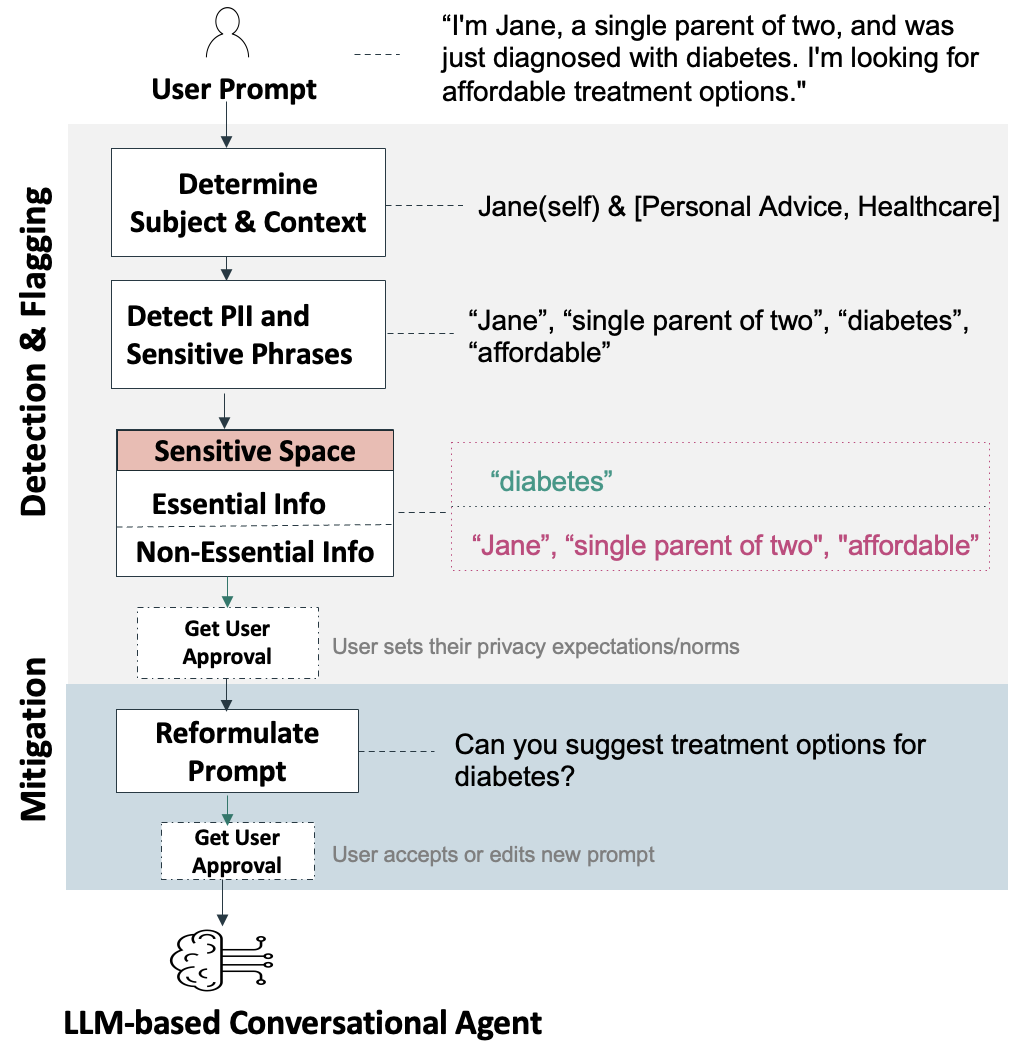}
    \caption{Overview of our framework for contextual privacy  in interactions with conversational agents. Our framework processes user prompts to identify context and sensitive information related to the context. It then provides reformulated prompts that maintain the original intent while reducing out-of-context information.}
    \label{fig:framework}
    \vspace{-9pt}
\end{figure}

LLM-based Conversational Agents (LCAs) such as chatbots, can offer valuable services to individual users \citep{mariani2023artificial,kumar2024medibot, yang2023ai, chow2023developing, rani2024ai, sadhu2024enhancing} in specialized systems such as customer service platforms and medical assistants, but  present unique privacy challenges that fundamentally differ from human-human interactions. For example, they can memorize \citep{carlini2019secret, biderman2024emergent, mccoy2023much, zhang2023counterfactual} and potentially misuse information \citep{kumar2024ethics}. They are vulnerable to data breaches or unauthorized sharing with third parties \citep{nagireddy2024socialstigmaqa,carlini2021extracting, nasr2023scalable}, and user-provided data may be incorporated into future model training, potentially resulting in unintended information leaks during deployment \citep{zanella2020analyzing}. 
In this paper, we focus on a critical but understudied aspect in user-LCA interactions: helping users make informed decisions about what information they share with these untrusted agents in the first place. This is particularly important because once information is shared with an LCA, users lose control over how it might be used or disseminated. Figure \ref{fig:framework} provides an overview of our proposed methodology to achieve this.

\paragraph{Motivation.} As LCAs become more adept at handling complex tasks and users remain uninformed about privacy risks, they develop increasing trust in both the technology and their own ability to protect themselves \citep{natarajan2020effects,cummings2023challenges}. 
Indeed, it has been shown that users are increasingly disclosing personal and sensitive information to LCAs \citep{zhang2024s, mireshghallah2024trust}.
In our own formative user study (Section~\ref{sec:framework}), we found that even expert participants are unaware of how indirect disclosures could reveal sensitive details in specific contexts. 
They expressed a desire for a real-time system that could highlight privacy risks and assist in revising information before sharing it with conversational agents. Similarly, our analysis of the real-world ShareGPT dataset \citep{vicuna2023}, reveals that users often share information beyond what their context requires, inadvertently exposing sensitive details that were unnecessary for their intended goals (see examples in  Table~\ref{tab:example_ci_violations}, details in Section~\ref{sec:framework}).

This motivates our main objective: %
\begin{center}
    \emph{Develop a framework that operates between users and conversational agents to detect and manage contextually inappropriate sensitive information during interactions.}
\end{center}

\paragraph{Contextual Privacy.} To enable the development of such a framework, we define the notion of \textit{contextual privacy} in user-LCA interactions, drawing ideas from the Contextual Integrity (CI) theory \citep{nissenbaum2004privacy, nissenbaum2011privacy}. Contextual integrity defines privacy not merely as hiding personal information, but as maintaining appropriate information flows within specific contexts. Drawing on the fundamental CI parameters, we define \textit{contextual privacy} by characterizing \texttt{User}$\rightarrow$\texttt{LCA} \textit{information flows} (Section~\ref{sec::threat_privacy}). Our contextual privacy notion requires that user prompts include only information that is contextually appropriate, relevant, and necessary to achieve the user's intended goals when interacting with LCAs, going beyond approaches that simply protect sensitive information~\citep{dou2023reducing,siyan2024papillon}. 
For instance, when a user is querying an LCA of a bank to locate tax forms, sharing SSN would adhere to contextual privacy, as it may be necessary for the task. On the other hand, if a user seeks advice on managing personal finances, sharing the names of family members would violate contextual privacy.

\paragraph{Proposed Framework.} We design a framework that can protect users during their interactions with LCAs. By analyzing user inputs, detecting potentially sensitive irrelevant content, and guiding users to reformulate prompts based on contextual relevance, our framework empowers users to make more informed, privacy-conscious decisions in real time. Rather than enforcing rigid privacy rules, the system helps users understand the privacy implications of their choices while preserving their intended interaction goals.

Our main contributions include:
\begin{itemize}[leftmargin=1em]
\setlength{\itemsep}{-2pt} 
    \item We formulate the definition of contextual privacy for the specific case of \texttt{User}$\rightarrow$\texttt{LCA} information flows, where users act as senders and LCAs as untrusted receivers;
    \item We apply our contextual privacy definition to analyze real-word conversation from ShareGPT \cite{vicuna2023} and demonstrate how users unintentionally violate contextual privacy in interactions with LCAs;
    \item We develop a privacy safeguarding framework that acts as an intermediary between the user and LCA, and helps users identify and reformulate out-of-context information in their prompts while maintaining their intended goals;
    \item We design novel metrics to measure the contextual privacy and utility performance of our framework;
    \item We show that our privacy safeguarding framework can be implemented using a small language models that can be locally deployed at the user side. We consider three state-of-the-art models for implementation, and compare their privacy and utility performances. Our experiments show that lightweight models can effectively implement this framework, achieving both strong privacy protection and utility preservation through different approaches to classify information relevant to the intended goals. 
    \item We conduct a human evaluation to assess the effectiveness of our contextual privacy framework and evaluation metrics. Our human evaluation showed that approximately 76\% of participants preferred the reformulated prompts over the original ones, supporting the framework’s practical utility and user trust. 
\end{itemize}

We fully contextualize our contributions with regards to existing literature in Appendix \ref{sec:relatedwork}.

\begin{table*}[t]
  \small
  \renewcommand{\arraystretch}{1.2}
  \centering
  \caption{Examples of contextual privacy violations in the ShareGPT dataset (determined by the judge model). Non-essential information that should be protected is highlighted in \textcolor{red}{red}, illustrating cases where unnecessary sensitive details were disclosed during interactions. Essential information is highlighted in \textcolor{blue}{blue}. }
  \label{tab:example_ci_violations}
  \vspace{1mm}
  \resizebox{0.94\textwidth}{!}{
  \begin{tabular}{p{2.5cm} p{11cm}}
    \toprule
    \textbf{User Intent} & \textbf{User Prompt} \\ 
    \toprule
    Looking for a job & \textcolor{red}{My friend Mark}, who was just \textcolor{red}{laid off from Google}, is \textcolor{blue}{looking for a job} where he \textcolor{blue}{can use ML and Python}. Do you have any advice for him? \\ 
    \midrule
    Pros and cons of running & I \textcolor{blue}{plan to go running at 18:30} today \textcolor{red}{with Gina and Emma} around \textcolor{blue}{île de la grande jatte in Levallois, France}. Give me the most likely negative outcome and the most likely positive outcome of this event. \\ 
    \midrule
    Cost of monthly medical checkup & \textcolor{red}{Wei's son} has \textcolor{red}{recently} been \textcolor{blue}{diagnosed with type 1 diabetes} which, according to him, will cost him an extra \textcolor{red}{\$200 per month}. How much extra will a monthly medical checkup cost? \\
    \midrule
     Write Poem & Please \textcolor{blue}{write a valentine's day themed poem} for \textcolor{red}{my wife Sandy}. Include our \textcolor{red}{13 week old daughter named Hailey} and add in some humor. \\
    \bottomrule
  \end{tabular}
  }
\end{table*}

\section{Threat Models and Privacy Definition}
\label{sec::threat_privacy}

\paragraph{Threat Model.} We consider a scenario where users interact with large, remote, and untrusted LCAs through APIs. These can be web-based or hosted on cloud-based services or private networks and may be either general-purpose or domain-specific. Users often share personal, financial, or medical information without clear knowledge of how their data is managed, increasing privacy risks due to the lack of transparency around these agents.

We focus on a threat model where users unintentionally compromise their privacy by oversharing information. Our approach targets out-of-context \emph{self-disclosure} by guiding users to share only contextually necessary information. By identifying unnecessary or sensitive disclosures in real time, we assist users in controlling the information they reveal, thereby reducing the risk of unintentional privacy breaches.
Our approach indirectly mitigates the threat of \emph{malicious users}, who seek to extract sensitive information from the agents by manipulating their interactions, by minimizing the amount of sensitive information exchanged during interactions.

\paragraph{Contextual Privacy in Conversational Agents.} We define the notion of \textit{contextual privacy} in conversational agents, inspired by the Contextual Integrity (CI) theory. 
CI models privacy as information flow defined by the five parameters sender (who is sharing the data), subject (who the information is about), receiver (who is getting the data), context (what sort of information is being shared), and transmission principle (the conditions under which information flow is conducted) \cite{nissenbaum2004privacy}. CI evaluates whether the information flow adheres to appropriate standards governed by norms, which vary based on the specific circumstances of the interaction. Establishing privacy norms and privacy principles of CI is complex and indeed an open problem in the literature since norms are governed by societal contexts and can evolve in response to societal developments \cite{malkin2023contextual}. 

Instead, we draw inspiration from the CI theory to formalize the notion of contextual privacy, focusing on the user-LCA interaction.
We begin with characterizing the \textit{information flow} between a user and an LCA by drawing on the five essential CI parameters in Table \ref{table:ci_params}. We simplify the transmission principle based on the privacy directive \textit{share information that is essential to get the answer}, similar to \cite{bagdasaryan2024air}.

After we characterize the subject and the \textit{context} (which captures the user’s intent and the key task) from the user’s query along with the prior conversation history, we determine two types of sensitive attributes in the query: (a) details that are essential to answer the query, and (b) sensitive details that are not essential for answering the query. We say that a user query is \textit{contextually private} if it does not contain any nonessential sensitive attributes. An example of essential and non-essential attributes for a query is shown in Figure \ref{fig:framework}.

\begin{table*}[t]
  \small
  \renewcommand{\arraystretch}{1.2}
  \centering
  \caption{Entities associated with contextual integrity in conversational agents.}
  \resizebox{0.94\textwidth}{!}{
  \begin{tabular}{ p{2cm}  p{4cm}  p{8cm} }
    \toprule
    \textbf{CI Entity} & \textbf{Definition} & \textbf{Function/Considerations} \\ 
    \toprule
    \textbf{Sender (self)} & The user sending information to the agent to achieve a task. & Ensure the user shares only relevant and necessary information. \\ 
    \midrule
    \textbf{Subject} & The individual(s) about whom information is shared (self, others, or both). & Protect the privacy of the subject by identifying whether the subject is the user or another person. Information shared should respect the subject's privacy. \\ 
    \midrule
    \textbf{Receiver (agent)} & The agent that receives and processes information. & Treat agent as untrusted. Apply strict privacy controls to prevent oversharing. May be domain-specific (e.g., MedicalChat Assistant) or general-purpose (e.g., ChatGPT). \\ 
    \midrule
    \textbf{Context (data type)} & The broader domain or user intent (e.g., medical, finance, work-related) guiding the interaction. & Guides what information is relevant to share. In domain-specific apps, the context is predefined; in general-purpose apps, intent detection is used. Optionally, users may specify sensitive contexts. \\ 
    \midrule
    \textbf{Transmission Principle} & The rule governing the flow of information between sender and receiver. & Share only essential and relevant information for the task, avoiding unnecessary or sensitive information. Respect the privacy expectations defined by context and actors. \\ 
    \bottomrule
  \end{tabular}}
  \label{table:ci_params}
\end{table*}

\vspace{-5pt}
\section{A Framework for Safeguarding Contextual Privacy}
\label{sec:framework}
\vspace{-5pt}

Our goal is to develop a framework that acts as an intermediary between the user and LCA, and enables the user to detect whether their prompt incurs any contextual privacy violations, and judiciously reformulate the prompt to ensure contextual privacy. We first conduct a formative design study to guide our framework design.

\paragraph{Formative User Study to Guide Our Framework Design.} We conducted a \textit{Wizard-of-Oz} formative user study to explore users' expectation of privacy when interacting with LCAs and to gather technical requirements for our framework.
Following established practices in early-stage interface design research \citep{ nielsen2000fiveusers, budiu2021fiveparticipants, nielsen1993mathematical}  where 5 participants are typically sufficient to identify major design insights, we conducted our study with six participants from our institution who were familiar with LLMs. Though small in scale, formative Human-Computer Interaction studies of this type have been shown to uncover the majority of usability issues~\citep{nielsen2000fiveusers} and provide rich design insights that complement large-scale evaluations~\citep{gaver1999design}.
Using three mid-fidelity UX mockups (see Appendix~\ref{sec::user_study_mockups}) as design probes~\citep{gaver1999design}, we explored how participants perceive, manage, and would prefer to be supported in protecting their privacy during LLM interactions. The mockups simulated real-world scenarios drawn from ShareGPT where both obvious and subtle privacy leaks could occur. Participants discussed how they currently think about privacy, what support they would value, and what tradeoffs they were willing to make.

Insights from this formative phase shaped several key design aspects of our framework, including distinguishing between essential and non-essential sensitive information, real-time feedback, user control over reformulations, and transparency around how sensitive information is handled and flagged. The participants rated the overall approach of the system highly, with a min and max rating of 7/10 and 9/10 respectively, providing initial validation for our approach to sensitive information detection and reformulation. For a detailed discussion of the study and how it impacted our design, see Appendix~\ref{sec::user_study}.
 
\paragraph{Proposed Framework.} We propose a framework that acts as an intermediary between the user and the conversation agent and enables the user to detect out-of-context sensitive information in the user prompt and judiciously reformulate the prompt to ensure contextual privacy.
The key components of the framework are outlined in Figure~\ref{fig:framework}.
When a user submits a prompt, our framework first determines the \textbf{context} and \textbf{subject} of the conversation. The context is divided into two components: the domain of the interaction (e.g., medical, legal, or financial) and the specific task the user aims to perform, such as seeking advice, requesting a translation, or summarizing a document. 
Context identification is guided by a taxonomy of common user tasks and sensitive contexts that go beyond PII (Personally Identifiable Information) \citep{mireshghallah2024trust} (see Appendix~\ref{domains_and_tasks}).

Once the context and subject are identified, our framework moves on to detecting sensitive information in the prompt. %
The framework categorizes the sensitive information into two spaces: (a) \textbf{essential information space}: sensitive details necessary to answer the user’s query, (b) \textbf{non-essential information space}: sensitive details that are unnecessary for answering the query and should be kept private.

In the example of Figure \ref{fig:framework}, the sensitive terms are \textit{``Jane'', ``single parent of two'', ``diabetes'', and ``affordable''}. While ``diabetes'' is essential for providing advice on treatment options, the other details---Jane's name, family situation, and financial concerns---are not required and thus classified as non-essential. 

Once contextually essential and non-essential information is identified, our framework improves contextual privacy by \textbf{reformulating} the prompt. This process includes removing, rephrasing, or redacting details within the non-essential information space, while preserving the user's intent. This way, we ensure that the user can still achieve the desired outcome effectively when the reformulated prompt is sent to the untrusted LCA. In our running example, a reformulated user's prompt could be \textit{``I need advice on managing a health condition and finding treatment options for diabetes''}, which protects non-essential sensitive details like the user’s name and personal circumstances, while maintaining the core intent of seeking treatment advice for diabetes.

After the reformulated prompt is generated, users can review, modify, or accept it, or revert to the original input. The review steps, shown by dashed boxes in Figure~\ref{fig:framework}, ensure user control, allowing them to achieve their desired balance between privacy and utility. The framework continues to highlight privacy implications as users adjust the suggested reformulation, helping them make informed choices about what information to share. Once finalized, the reformulated prompt is sent to the LLM-based conversational agent to obtain a response.

\section{Implementation and Evaluation}

\begin{figure*}[htbp]
\centering
    \includegraphics[width=1.0\linewidth,keepaspectratio]{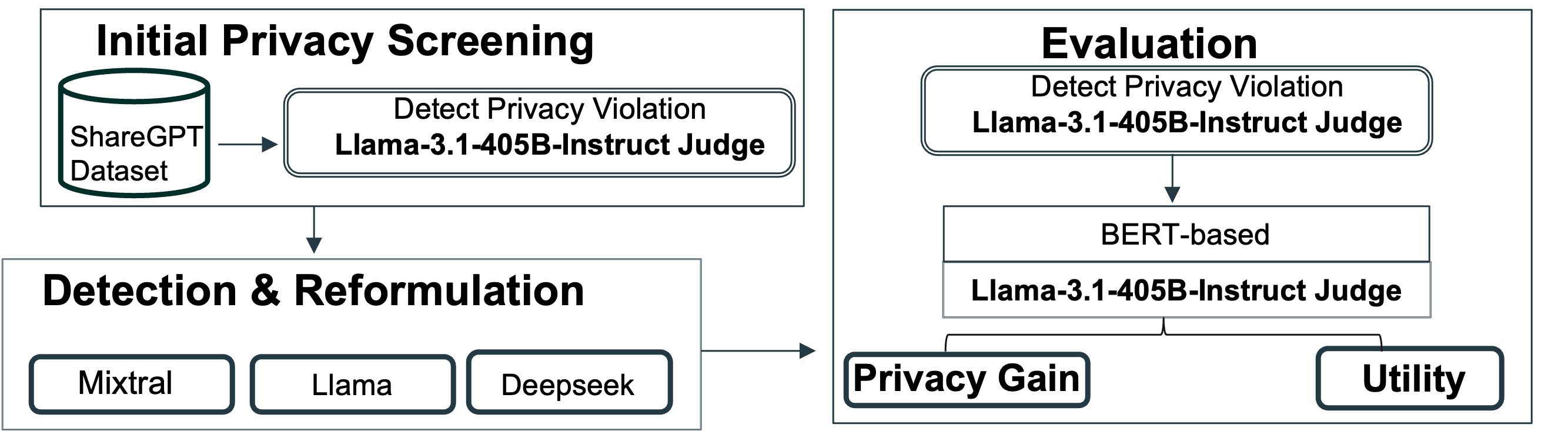}
    \caption{Experimental pipeline showing initial privacy screening, reformulation by three local models, and evaluation stages.}
    \label{fig:experiment}
\end{figure*}

\subsection{Contextual Privacy Evaluation of Real-World Queries}
\label{sec:sharegot_privacy_evaluation}
Before implementing and evaluating our framework, we first perform initial privacy analysis by evaluating 
an open-source version of the ShareGPT dataset~\citep{vicuna2023} to understand the prevalence of contextual privacy violations. To instantiate our formal privacy definition, we used Llama-3.1-405B-Instruct \cite{grattafiori2024llama3herdmodels} as judge, with a prompt designed to identify violations of contextual integrity (Appendix \ref{appendix_ci_detection}). From over 90,000 conversations, we retain 11,305 single-turn conversations within a reasonable length range (25-2,500 words). For each conversation, the judge model assessed the context, sensitive information, and their necessity for task completion. This analysis identified approximately 8,000 conversations containing potential contextual integrity violations. To manage inference costs, we focused on cases where the judge model could successfully identify a primary context and classify essential and non-essential information attributes, yielding 2,849 conversations (25.2\%) with definitive contextual privacy violations. Examples of these violations are shown in Table \ref{tab:example_ci_violations}. Manual inspection of the judge's results for consistency and correctness demonstrated good classification performance with few false positives and negatives.

\subsection{Implementation Details}
\paragraph{Models.} We implement our framework using a model that is significantly smaller than typical chat agents like ChatGPT, enabling users to deploy the model locally via Ollama\footnote{\url{https://github.com/ollama/ollama}} without relying on external APIs.
In our experiments, we evaluate three models with different characteristics: Mixtral-8x7B-Instruct-v0.1\footnote{\url{https://ollama.com/library/mixtral:8x7b-instruct-v0.1-q4\_0}} \cite{jiang2024mixtralexperts}, Llama-3.1-8B-Instruct\footnote{\url{https://ollama.com/library/llama3.1:8b-instruct-fp16}} \cite{grattafiori2024llama3herdmodels}, 
and DeepSeek-R1-Distill-Llama-8B\footnote{\url{https://ollama.com/library/deepseek-r1:8b-llama-distill-q4_K_M}} (focused on reasoning) \cite{deepseekai2025deepseekr1}. We refer to these models as Mixtral, Llama and Deepseek in short going forward.
The local deployment of models ensures no further privacy leakage due to the framework. Although our evaluation focuses on three LLMs, our approach is model-agnostic and can be applied to other architectures. For assessment of privacy and utility, we use Llama-3.1-405B-Instruct \cite{grattafiori2024llama3herdmodels} as an impartial judge, which was hosted in a secure cloud infrastructure.

\paragraph{Experiment Setup.} 
As discribed in the previous section,
our framework processes user prompts in three stages: (a) context identification, (b) sensitive information classification, and (c) reformulation. The locally deployed model first determines the context of the conversation, identifying its domain and task (Appendix ~\ref{domains_and_tasks}) using the prompts in Appendix Appendix~\ref{appendix_intent_detection} and Appendix ~\ref{appendix_task_detection} respectively. It then detects sensitive information, categorizing it as either \textit{essential} (required for task completion) or \textit{non-essential} (privacy-sensitive and removable). Finally, if non-essential sensitive information is present, the model reformulates the prompt to improve privacy while preserving intent.

We implement two approaches for sensitive information classification: \textbf{dynamic classification}  and \textbf{structured classification}, each reflecting different ways to operationalize our privacy framework. In the \textbf{dynamic classification approach} (see prompt used in Appendix ~\ref{appendix_dynamic_sentive_template}),  the model determines which details are essential based on how they are used within the specific conversation. For instance, in the prompt \emph{"I’m Jane, a single parent of two, and was just diagnosed with diabetes. I’m looking for affordable treatment options"}, the model would identify the phrases= \emph{["diabetes"]} as the essential attributes, while \emph{["Jane", "single parent of two","affordable"]} would be classified as non-essential. This adaptive method aligns with contextual privacy formulation, ensuring that only task-relevant details are retained. In contrast, the \textbf{structured classification approach} (see prompt used in Appendix ~\ref{appendix_structured_sentive_template}), allows to specify a predefined list of sensitive attributes (e.g., age, SSN, physical health, allergies) that should always be considered non-essential (protected), ensuring consistent enforcement of privacy policies.  For the same example, this approach would flag \emph{["physical health"]} as the essential attribute while labeling \emph{["name", "family status", "financial condition"]} as non-essential attributes, recommending them for removal based on user-defined privacy preferences. This provides greater control over what information is considered sensitive, allowing customization while maintaining a standardized privacy framework. The predefined attribute categories follow those defined in \citet{bagdasaryan2024air}.

If non-essential sensitive details are detected, the model reformulates the prompt by either removing or rewording them to minimize privacy risks while maintaining usability (see Prompt used in Appendix \ref{appendix_reformulation}). By evaluating both dynamic and structured classification, we demonstrate the flexibility of our framework in balancing adaptability with user-defined privacy controls.

\subsection{Evaluation and Results}

We evaluate our framework by measuring two key metrics: \textbf{privacy gain} and \textbf{utility}. Privacy gain quantifies how effectively sensitive information is removed during reformulation, while utility measures how well the reformulated prompt maintains the original prompt's intent. We compute these metrics using two complementary methods: an automated BERTScore-based comparison of sensitive attributes, and an LLM-based assessment that aggregates multiple evaluation aspects.

\subsubsection{Evaluation via Attribute-based Metrics} 
\paragraph{Metrics.} 
We measure privacy gain by computing semantic similarity between non-essential attributes between original and reformulated prompts, where similarity is computed using BERTScore \citep{zhang2020bertscore}. 
Specifically, we first run the judge model on reformulated prompts to obtain non-essential sensitive attributes $\mathcal{P}^{\textrm{reform}}_{\textrm{non-ess}}$, using a prompt designed to identify contextual privacy violations (Appendix \ref{appendix_ci_detection}). We have non-essential sensitive attributes for original prompts $\mathcal{P}^{\textrm{orig}}_{\textrm{non-ess}}$ from Section \ref{sec:sharegot_privacy_evaluation}. 
Given sets of strings $\mathcal{P}^{\textrm{orig}}_{\textrm{non-ess}}$ and $\mathcal{P}^{\textrm{reform}}_{\textrm{non-ess}}$, 
privacy gain is computed as
$1 - \text{BERTScore}(\mathcal{P}^{\textrm{orig}}_{\textrm{non-ess}}, \mathcal{P}^{\textrm{reform}}_{\textrm{non-ess}})$, with a score of 1.0 assigned when either set is empty. A higher privacy gain indicates better removal of sensitive information. For utility, we measure semantic similarity between essential attributes using $\text{BERTScore}(\mathcal{P}^{\textrm{orig}}_{\textrm{ess}}, \mathcal{P}^{\textrm{reform}}_{\textrm{ess}})$, where a score closer to 1.0 indicates better preservation of task-critical information. Since BERTScore works on text pairs, we match each original attribute to its closest reformulated one and compute utility as the fraction of matched attributes above a similarity threshold of 0.5.

\begin{table}[t]
  \small
  \setlength{\tabcolsep}{4pt}
  \centering
  \caption{BERT-based Evaluation of Privacy Gain and Utility}
  \begin{tabular}{lcc}
    \toprule
    \multicolumn{3}{c}{\textbf{Dynamic Attribute Classification}} \\ \midrule
    \textbf{Model} & \textbf{Privacy Gain $\uparrow$} & \textbf{Utility(BERTScore)$\uparrow$} \\ \midrule
    Deepseek  & 0.853 & 0.570 \\
    Llama     & 0.886 & 0.567 \\
    Mixtral   & 0.873 & 0.570 \\ \midrule
    \multicolumn{3}{c}{\textbf{Structured Attribute Classification}} \\ \midrule
    \textbf{Model} & \textbf{Privacy Gain $\uparrow$} & \textbf{Utility(BERTScore)$\uparrow$} \\ \midrule
    Deepseek  & 0.836 & 0.511 \\
    Llama     & 0.873 & 0.606 \\
    Mixtral   & 0.824 & 0.576 \\ \bottomrule
  \end{tabular}
  \label{tab:privacy_utility}
\end{table}

\paragraph{Results.} Table~\ref{tab:privacy_utility} shows that under dynamic classification, all three models achieve strong privacy scores (0.85-0.88) with comparable utility ($\sim0.57$), suggesting that the ability to identify context-specific sensitive information is robust across different model architectures.

The structured classification approach shows greater variation between models. While Llama achieves high scores in both privacy (0.873) and utility (0.606), structured classification generally yields slightly lower privacy scores but more variable utility. This suggests a natural trade-off: predefined categories might miss some context-specific sensitive information, yet operating within these fixed boundaries can help preserve task-relevant content. Interestingly, the similar performance patterns across different model architectures suggest that the choice between instruction-tuned and reasoning-focused approaches may be less crucial for privacy-preserving reformulation.

The success of both dynamic and structured approaches offers implementation flexibility - users can choose predefined privacy rules or context-specific protection based on their requirements. This choice, rather than model architecture, appears to be the key decision factor in deployment.

Since BERTScore-based evaluations utilize the semantic similarity between sensitive attributes, they do not consider situations in the reformulated query where some of these sensitive attributes may be hallucinated, merged, or split. We believe that these factors primarily contributed to the decrease in the utility. In the following section, we will explore using LLM-as-a-judge to address some of these challenges.

\begin{table}[t]
\small
\centering
\caption{LLM-as-a-Judge Evaluation of Privacy Gain and Utility}
\begin{tabular}{lcc}
\toprule
\multicolumn{3}{c}{\textbf{Dynamic Attribute Classification}} \\
\midrule
\textbf{Model} & \textbf{Privacy Gain $\uparrow$} & \textbf{Utility Score$\uparrow$} \\  
\midrule
Deepseek  & 0.802 & 0.845 \\ 
Llama     & 0.858 & 0.861 \\ 
Mixtral   & 0.848 & 0.838 \\ 
\midrule
\multicolumn{3}{c}{\textbf{Structured Attribute Classification}} \\ 
\midrule
\textbf{Model} & \textbf{Privacy Gain $\uparrow$} & \textbf{Utility Score$\uparrow$} \\  
\midrule
Deepseek  & 0.815 & 0.825 \\ 
Llama     & 0.855 & 0.858 \\ 
Mixtral   & 0.845 & 0.828 \\ 
\bottomrule
\end{tabular}
\label{tab:llm_judge_results}
\end{table}
\subsubsection{LLM-as-a-Judge Assessment} 
\paragraph{Setup.}  We use Llama-3.1-405B-Instruct as a judge to provide a complementary evaluation of privacy and utility across 100 randomly selected queries per model (6×100 total). Given the high computational cost of LLM-based inference, this targeted sampling allows us to validate key trends observed in the attribute-based evaluation while minimizing overhead. 
Privacy gain is computed by asking the judge to evaluate privacy leakage, coverage, and retention, while utility 
is computed by measuring
query relevance, response validity, and cross-relevance. 
These binary evaluations are averaged to produce final privacy gains and utility scores. See Appendix \ref{appendix_evaluation} for detailed prompts and evaluation criteria.

\paragraph{Results.} The LLM-based assessment shows generally higher utility scores (0.82-0.86) across all models compared to BERTScore-based evaluation, while maintaining similar privacy levels (0.80-0.86). This difference can be attributed to how attributes are detected and compared—BERTScore evaluates exact semantic matches between attributes, while the LLM judge takes a more holistic view of information preservation. For instance, when essential information is restructured (e.g., ``my friend Mark'' split into separate attributes), BERTScore may indicate lower utility despite semantic equivalence.

The LLM evaluation confirms the effectiveness of both classification approaches, with dynamic classification showing slightly more consistent performance across models. Llama maintains its strong performance under both approaches (privacy gain: $\sim0.85$, utility score: $\sim0.86$), reinforcing its reliability for privacy-preserving reformulation.

\subsubsection{Example Reformulations and Trade-offs}
\paragraph{Setup.} Table~\ref{tab:privacy_utility_scores} presents a set of diverse example reformulations illustrating our framework’s ability to balance privacy and utility across different scenarios. These examples highlight both ideal cases—where reformulation effectively preserves both privacy and utility—and more challenging ones where trade-offs are unavoidable.

\begin{table*}[t]
  \small
  \scriptsize
\renewcommand{\arraystretch}{1.7}
  \centering
  \caption{Examples of original and reformulated prompts with privacy and utility scores. These examples illustrate how our framework removes sensitive details while preserving task intent. The high privacy, high utility cases demonstrate successful reformulation, while the low privacy, low utility example highlights cases where effective reformulation is difficult or unnecessary.}
  \label{tab:privacy_utility_scores}
  \resizebox{0.99\textwidth}{!}{
  \begin{tabular}{p{5.5cm}p{5.5cm}cc}
    \toprule
    \textbf{Original Prompt} & \textbf{Reformulated Prompt} & \textbf{Privacy Gain} & \textbf{Utility (BERTScore)} \\ \hline
    My friend Mark who was just laid off from Google, is looking for a job where he can use ML and Python. Do you have any advice for him? 
    & Someone is looking for a job where they can use ML and Python skills. Do you have any advice? 
    & \cellcolor{green!20} \textbf{1.0} 
    & \cellcolor{green!20} \textbf{1.0} \\ \hline

    Please write a poem for my wife for Valentine's Day. Her name is Gili. Please use rhymes. Please make it sarcastic, but not too much so. A bit funny but romantic.  
    & Write a poem for Gili, incorporating sarcasm at an appropriate level and maintaining a tone that is both funny and romantic  
    & \cellcolor{green!20} \textbf{1.0} 
    & \cellcolor{yellow!20} \textbf{0.5} \\ \hline

     I am writing a character letter to a judge in hopes of having charges against my friend Amber to be dropped. Write a letter in a positive tone to a judge speaking to Amber's good character ...... How you know me, my ties to the community my profession and stating a dismissal is imperative to my earning potential ... I know amber from "Teach For America". I know her to be a passionate educator who cares about uplifting the youth and her community. She should have the charges dropped so that she can increase her community engagement and earning potential for her family.  
    & Write a character letter in a positive tone to a judge speaking to someone's good character, highlighting their community ties and profession, and stating that a dismissal is imperative to their earning potential. I know this person from a professional program and they have been a passionate advocate for the betterment of their community. The charges should be dropped so that they can increase their engagement in the community and improve their financial situation. 
    & \cellcolor{yellow!20} \textbf{0.5} 
    & \cellcolor{green!10} \textbf{0.83} \\ \hline

    Sunny Balwani : I worked for 6 years day and night to help you. Elizabeth Holmes : I was just thinking about texting you in that minute by the way  
    & Sunny Balwani : I am responsible for everything at Theranos. Elizabeth Holmes : .........  
    & \cellcolor{red!20} \textbf{0.0} 
    & \cellcolor{red!20} \textbf{0.0} \\ 
    \bottomrule
  \end{tabular}
  }
\end{table*}

\paragraph{Results.} Our framework successfully removes personal identifiers while preserving task relevance, as seen in the third example ($\text{privacy gain} = 0.5,\text{utility score}= 0.83$). In creative requests like the Valentine’s poem (second example), removing personal details reduces privacy risks but slightly impacts personalization $\text{utility}=0.5$).

Some contexts resist reformulation. The last conversation ($\text{privacy gain}= 0.0, \text{utility score}= 0.0$) highlights cases where the entire prompt is inherently sensitive, requiring alternative privacy measures beyond text transformation. This is not a failure of our approach but an indicator of when reformulation alone is insufficient.

These examples reinforce that privacy-preserving reformulation is a trade-off, not a one-size-fits-all solution.

\subsection{Comparison with Other Baselines}
To contextualize the performance of our framework, we compared it against Microsoft’s Presidio~\citep{presidio}, a widely used tool for redacting PII-based on pattern matching. While Presidio is effective at identifying explicit PII (e.g., names, emails), it does not assess whether information is appropriate to share based on the context or user intent.

We conducted an evaluation of both systems using 20 ShareGPT prompts exhibiting contextual privacy risks. Using our LLM-as-a-judge evaluation, our framework achieved a perfect average utility score of \textbf{1.0}, compared to Presidio’s \textbf{0.89}. Both the systems achieved comparable privacy scores of around \textbf{0.7}. This demonstrates that our approach better preserves contextually essential information, offering a stronger utility-privacy balance than pure PII redaction.

We also considered AirGap~\citep{bagdasaryan2024air}, which uses contextual integrity to limit what AI assistants reveal to third parties. In contrast, our work guides users during interactions to reduce oversharing. AirGap assumes persistent access to user data and acts independently of the user; our approach is local, user-facing, and operates in real-time.  These approaches serve different goals—preventing user oversharing versus limiting agent disclosures—and rely on distinct workflows and evaluation setups. A direct comparison is therefore not meaningful.

\section{Human Evaluation Results}

We conducted a human evaluation with 24 participants to assess the effectiveness of our contextual privacy framework. Each participant reviewed a set of 20 prompts from ShareGPT, along with their corresponding reformulated versions produced by our system.

Participants answered multiple-choice questions to evaluate whether the reformulated prompts successfully removed non-essential sensitive information, retained essential information, preserved the original user intent, remained natural and coherent, and were preferred over the original queries. They were also asked whether they agreed with the classifications made by the LLM-as-a-judge. The results show strong support for the framework:

\begin{itemize}[leftmargin=1em,itemsep=2pt,parsep=0pt]
    \item 76.2\% of participants preferred the reformulated prompts over the original ones.
    \item 100\% agreed that non-essential sensitive information was effectively removed.
    \item 85.7\% stated that essential information was retained.
    \item 95.2\% agreed that the reformulations preserved the original intent.
    \item 100\% found the reformulated prompts natural and coherent.

\end{itemize}

Based on our observation from the human evaluation, we acknowledge that contextual privacy is quite subjective and the relationship between privacy preservation and task effectiveness can vary significantly across different contexts, cultural variations, and user preferences. The complete set of survey questions is included in Appendix~\ref{appendix_human_eval}.

\subsection{Human Assessment of Evaluation Metrics (LLM-judge \& BERTScore) }
To assess the reliability of our automated metrics, we examined their correlation with human judgments. We found that 90.5\% of participants agreed with most or all of the classifications of essential and non-essential sensitive information used in our evaluation metrics, suggesting strong alignment between our automated evaluator and human perspectives. Further correlation analysis yielded statistically significant results: Pearson’s $r=0.55$ ($p=0.0095$), Spearman’s $\rho=0.49$ ($p=0.02$), and Kendall’s $\tau=0.47$ ($p=0.02$). These findings show that our utility and privacy scores closely reflect human judgment, allowing us to evaluate the system at scale without needing manual annotation.

\section{Discussion and Conclusion}
\vspace{-5pt}
Drawing ideas from the contextual integrity theory, we defined the notion of contextual privacy for users interacting with LLM-based conversation agents.
We proposed a framework, grounded in our contextual privacy formulation, that acts as an intermediary between the user and the agent, and carefully reformulates user prompts to preserve contextual privacy while preserving the utility.

This work serves as an initial step in exploring privacy protection in user interactions with conversational agents. There are several directions that future research can further investigate. 
First, our framework may not be suitable for user prompts that require preserving exact content, such as document translation or verbatim summarization. For example, translating a legal document demands keeping the original content intact, making it challenging to reformulate while preserving contextual privacy. For such tasks, alternative approaches like using placeholders or pseudonyms for sensitive information could help protect privacy without compromising accuracy, though this is beyond our current implementation. 
Second, our framework relies on LLM-based assessment of privacy violations which, while effective for demonstrating the approach, lacks formal privacy guarantees and can be sensitive to the prompt. Future work could explore combining our contextual approach with deterministic rules or provable privacy properties. 
Third, while we demonstrate how users can adjust reformulations to balance privacy and utility, developing precise metrics to quantify this trade-off remains an open research challenge. This is particularly important as the relationship between privacy preservation and task effectiveness can vary significantly across different contexts and user preferences. 

Finally, while our evaluation using selected ShareGPT conversations demonstrates the potential of our approach, broader testing across diverse contexts and user groups would better establish the framework's general applicability.

\section*{Limitations}
\vspace{-4pt}
Contextual integrity is a relatively new and fluid notion of privacy. Ours is also one of the very early works exploring this space from the standpoint of LLM-based conversational agents. Naturally, this leads to a number of challenges, some of which are beyond the scope of the work and should be addressed in the future. Like we discussed before, establishing privacy norms and principles in CI itself is complex and dependent on societal contexts, which is why we restrict ourselves to a practical and useful variation of the idea. However, developing templates for implementing CI under various societal contexts deserves significant attention from the research community in the future.

Our framework addresses critical privacy concerns in LLM interactions, potentially shaping future norms around data sharing in conversational AI. By enhancing user awareness and control over sensitive information, it promotes more ethical AI deployments, safeguarding user privacy in diverse applications such as healthcare, legal, and personal assistance. However, there are ethical challenges, such as ensuring fairness across cultural contexts and preventing over-reliance on automated privacy detection. 

Additionally, our current implementation is limited to English-language prompts and focuses primarily on single-turn interactions. While the framework can support multi-turn conversations, we do not yet examine how reformulations affect evolving context, user experience, or cumulative privacy exposure over time. Extending the approach to multilingual and multi-turn settings, where privacy risks may compound, remains important future work. The pipeline also assumes access to moderately-sized LLMs, which may limit deployment in resource-constrained environments. 

The proposed framework natively supports multi-turn conversations. We can apply our framework iteratively for each interaction with the LLM with the history of the conversation as a part of the context. However, the key limitation in the current framework is that it doesn’t support conversions where the user intent can drift. For example, a conversation about `mental health’ could turn into a conversation about `financial condition’ intent.
Finally, although our LLM-as-a-judge metrics align well with human assessments, their broader validity across varied interaction types and user populations requires further investigation.

\section*{Acknowledgments}
We thank Shubhi Asthana, Chad DeLuca, Erik Miehling, Jessica He, Stephanie Houde, and Anupama Murthi for their helpful feedback, discussions, and support throughout this project. Their contributions were instrumental in refining the ideas and improving the quality of this work.

\bibliography{custom}

\appendix

\section{Related Work}
\label{sec:relatedwork}
We fully contextualize our contributions in regard to existing literature here.

\paragraph{LLM Privacy-Preserving Techniques.}

A significant body of research on privacy preservation in LLMs has focused on the training phase \citep{zhang2024dpzero,chua2024mind,yu2021differentially,yue2022synthetic,li2021large}. Techniques like differential privacy (DP)~\citep{dwork2006calibrating} have been used to prevent LLMs from memorizing sensitive information during training. Additionally, data sanitization strategies, such as deduplication and anonymization, have been used to reduce privacy risks by removing sensitive data from training data~\citep{lison2021anonymisation, kandpal2022deduplicating}. After training, machine unlearning methods have emerged to help eliminate any retained private data~\citep{carlini2019secret, biderman2024emergent, mccoy2023much, zhang2023counterfactual, carlini2021extracting, nasr2023scalable, xu2024machine}. However, inference-phase privacy protection has received less attention, with limited approaches, such as PII detection and DP decoding, targeting the risks of exposing sensitive information in real-time interactions with LLMs~\citep{majmudar2022differentially,carey2024dp,wu2023privacy,tang2023privacy,hong2023dp,edemacu2024privacy}. Recently, \citet{mireshghallah2023can} highlighted this gap, showing that LLMs often fail to protect private information in context and emphasizing the need for better privacy-preserving techniques. Our approach addresses this need by offering real-time, context-aware privacy guidance during user interactions, allowing individuals to better manage what information they disclose during conversations with LLMs.

\paragraph{Privacy Risks in Human-LLM Interactions.}
Self-disclosure during human-machine interactions can result in unintended sharing of sensitive information. For example, \citet{ravichander2018empirical} found that users tend to reciprocate with automated systems, revealing more personal information over time. Building on this, \citet{zhang2024s} examined the privacy risks faced by users interacting with LLMs, showing that human-like responses can encourage sensitive disclosures, complicating privacy management. \citet{mireshghallah2024trust} further advanced this discussion by highlighting the limitations of PII detection systems, showing that users often disclose sensitive information that goes beyond PII~\citep{cummings2023challenges, dou2023reducing}. Our work builds on these efforts by showing that users frequently disclose unnecessary information during interactions with LLMs, which can be contextually sensitive and unrelated to their intended goals. We develop a system that detects such information and offers reformulation suggestions to guide users toward more privacy-aware interactions.

\paragraph{Data Minimization in ML.}
The principle of data minimization, central to privacy regulations like GDPR~\citep{voigt2017eu}, has recently been a key focus in ML research. For example, \citet{ganesh2024data} formalized data minimization within an optimization framework for reducing data collection while maintaining model performance. \citet{tran2024data} expanded on this by showing that individuals can disclose only a small subset of their features without compromising accuracy, thus minimizing the risk of data leakage. While both approaches focus on reducing the amount of data processed during inference, our work applies data minimization in real time, guiding users to share only necessary information with LLMs. We integrate CI to ensure that the disclosed information aligns with the context of the conversation, ensuring GDPR compliance through a user-driven, context-aware approach.
\paragraph{Operationalizing Contextual Integrity (CI).}
Research on contextual privacy in LLMs is rapidly expanding. For instance, \citet{mireshghallah2023can} introduced a benchmark to evaluate the privacy reasoning abilities of LLMs at varying levels of complexity, while \citet{shvartzshnaider2024llm} proposed a comprehensive framework using CI to assess privacy norms encoded in LLMs across different models and datasets. CI has also been integrated into various practical systems to safeguard privacy across diverse domains. For example, \citet{shvartzshnaider2019vaccine} employed CI to detect privacy leaks in email communications, and \citet{kumar2020aquilis} applied CI to provide mobile users with real-time privacy risk alerts. In smart home ecosystems, \citet{malkin2022runtime,abdi2021privacy} used CI to analyze and enforce privacy norms. \citet{hartmann2024can} considered scenarios where a local model queries a larger remote model, leveraging CI to ensure only task-relevant data is shared. Similarly, \citet{bagdasaryan2024air} used CI to restrict AI assistants’ access to only the information necessary for a given task, and \citet{ghalebikesabi2024operationalizing} applied CI to ensure form-filling assistants follow contextual privacy norms when sharing user information. These approaches apply CI from the agent’s perspective, controlling what AI systems share with third parties. In contrast, our work focuses on the user-to-LLM interaction, helping users avoid oversharing with the agent itself. While AirGap addresses adversarial context hijacking by limiting agent disclosures, we target inadvertent user disclosures by providing real-time reformulation guidance. AirGap works passively without user input, whereas our system actively involves the user in the loop. These differences make the two approaches complementary rather than directly comparable. Also, while these studies focus on aligning AI assistants' actions with privacy norms, our work shifts the perspective toward empowering privacy-conscious users. By integrating CI into our framework, we aim to educate users in real time about contextually sensitive disclosures and offer proactive guidance to help manage privacy risks. This user-centered approach not only protects sensitive information during AI interactions but also promotes long-term privacy awareness---an aspect often overlooked in system-oriented solutions.

\section{Formative User Study to Guide System Design}
\label{sec::user_study}

To explore users' perceptions of privacy with LCAs and gather technical requirements for our framework, we conducted a Wizard-of-Oz formative user study with six participants from our institution who were generally familiar with LLMs. 

The study involved a 30-minute semi-structured interview where participants were presented with three mid-fidelity UX mockups, each designed to demonstrate different ways private and sensitive information could be detected and remediated (see Appendix~\ref{sec::user_study_mockups}). These mockups, featuring synthetic examples inspired by real-world patterns in the ShareGPT dataset, were created to expose participants to targeted privacy risks, such as unintentional PII and sensitive data disclosures. We used these mockups to probe participants' views on their own privacy practices, their thoughts about privacy disclosures, and their preferences for managing sensitive information in conversations. The study provided insights into people's views on the identification, flagging, and reformulation of sensitive data, shaping the core elements of our framework.

\begin{itemize}[leftmargin=1em]
    \item \textbf{Perceived privacy control}. Participants initially believed their efforts to protect their privacy when using real-world LLM applications were effective due to how they kept conversations vague. After they saw real examples of indirect privacy leaks in the mockups, many participants expressed greater concern about unintentionally sharing private information. \textbf{Design impact}: This insight emphasized the importance of identifying both direct and indirect privacy risks during LLM interactions in our system.
    \item \textbf{Visual identification of sensitive information}. Prototype B's color-coded differentiation between PII, necessary, and unnecessary information was praised for making privacy risks clearer and easier to understand. \textbf{Design impact}: Based on this feedback, we included the ability to differentiate between different kinds of sensitive information disclosures to help inform users' decision-making.
    \item \textbf{Reformulation preferences}. Although some participants preferred doing the work of reformulating their LLM prompts themselves, most wanted the system to offer (at least) one reformulated prompt suggestion, with the option to generate new suggestions. A few participants suggested offering multiple reformulations at once, selected across a spectrum of privacy-utility tradeoffs. In this way, users can balance their level of privacy protection with the utility of the output. \textbf{Design impact}: We designed our system to present one reformulation recommendation at a time, but with the flexibility to generate new alternative reformulations. In future iterations of our system, we plan to explore how to generate multiple reformulation options across varied privacy-utility tradeoffs.
    \item \textbf{User control and real-time feedback}. Real-time feedback and user control over editing flagged prompts were highly valued. Participants preferred having the system automatically generate reformulations, but they wanted the ability to make any necessary final adjustments. \textbf{Design impact}: We implemented a review step where users can edit, accept, or proceed with the original input before final submission to the LLM, providing the flexibility users requested.

    \item \textbf{Positive reception}. Participants responded positively to the system’s potential for managing sensitive information, with an average rating of $8.7 (\pm 0.87)$ on the importance of detecting and flagging sensitive details. \textbf{Design impact}. This feedback reinforced the central role of sensitive information detection in our framework, highlighting its perceived value to users.
    \item \textbf{Clarity and transparency}. Participants expressed a strong desire for transparency about how the system operates, including which tools or models are being used, and the meaning of key terms like ``necessary'' versus ``unnecessary'' information. \textbf{Design impact}: Our framework ensures transparency by detailing how sensitive information is identified and handled, including the models used, how they are applied, deployed, and how data is managed. We recommend real-world implementations do the same to build user trust.
    \item \textbf{Broader application}. A few participants suggested applying the tool to other contexts beyond LLM chat interfaces, such as search engines. \textbf{Design impact}:  This feedback highlights the importance of managing sensitive information and the broader applicability of our approach to other contexts.
\end{itemize}

\subsection{Formative User Study Design Mockups}
\label{sec::user_study_mockups}
\begin{figure*}[ht]
    \centering
    
    \begin{minipage}{0.8\textwidth} %
        \centering
        \includegraphics[width=\linewidth]{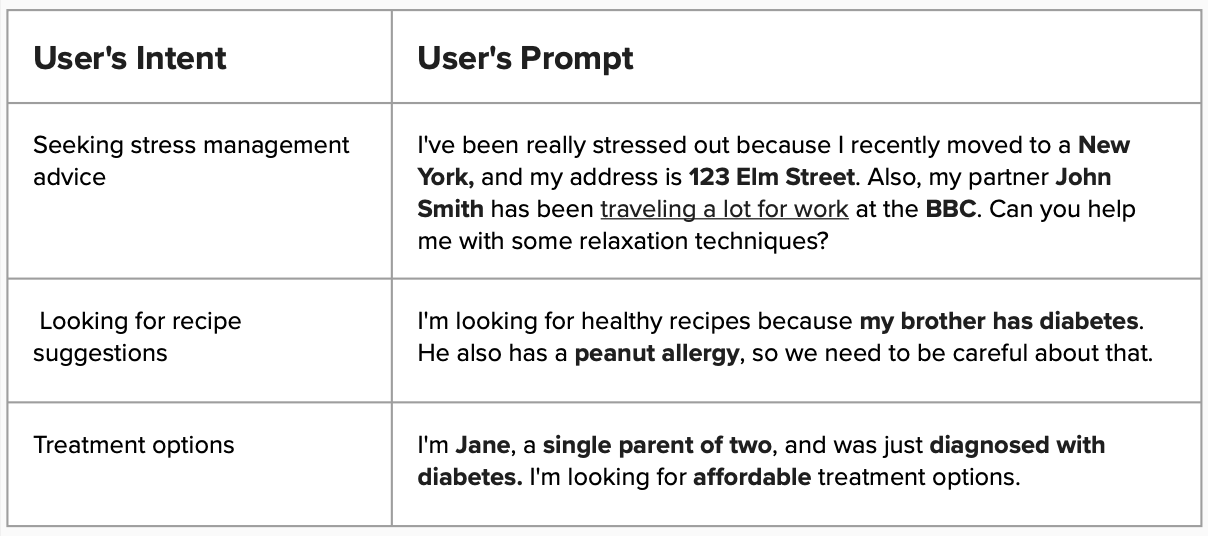}
        \subcaption{Examples of unintentional disclosures shown to participants}
    \end{minipage}

    \vspace{0.5cm} %

    \begin{minipage}{0.45\textwidth}
        \centering
        \includegraphics[width=\linewidth]{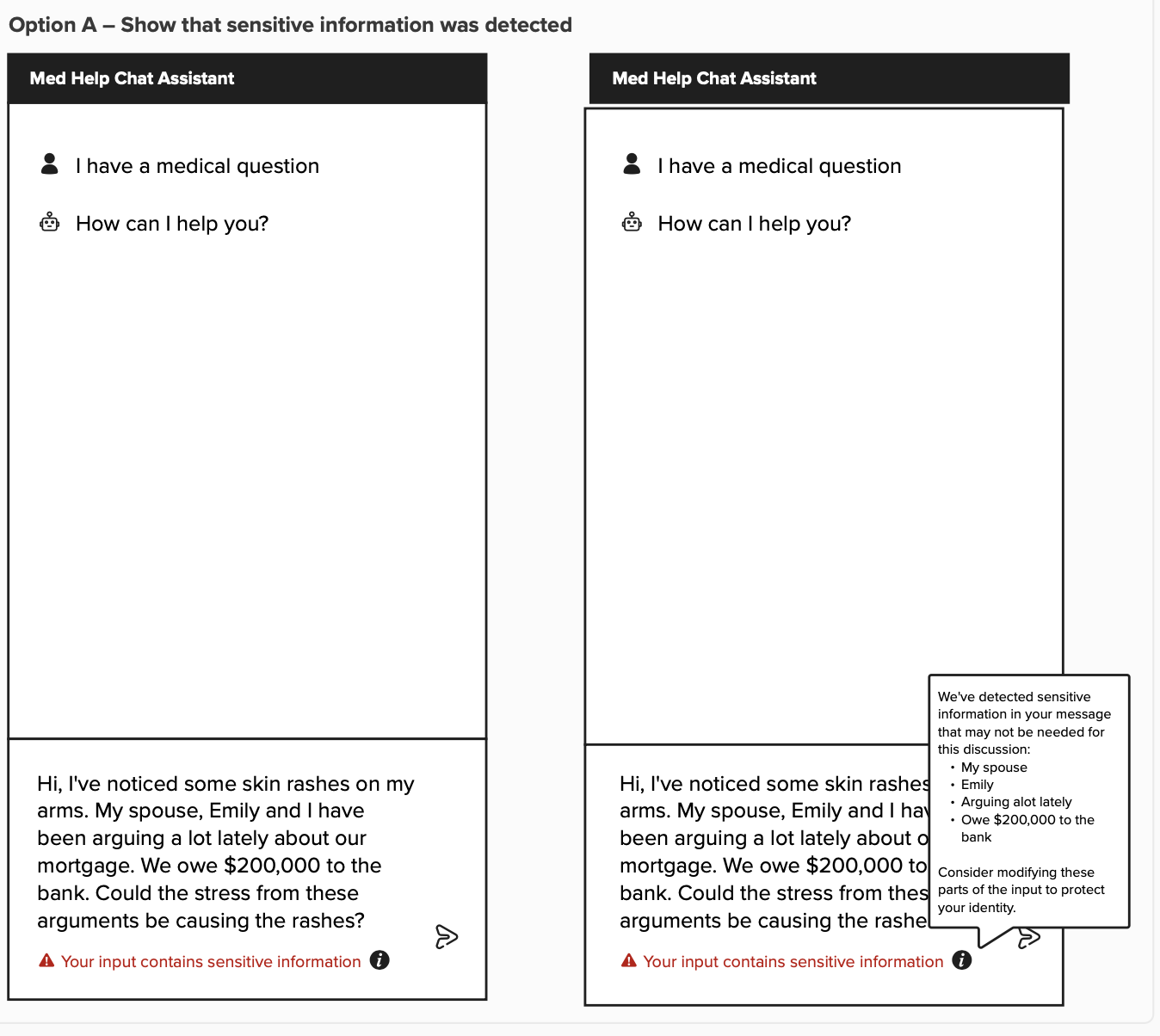}
        \subcaption{Mockup 1: Display all detected sensitive info}
    \end{minipage}
    \hfill
    \begin{minipage}{0.52\textwidth}
        \centering
        \includegraphics[width=\linewidth]{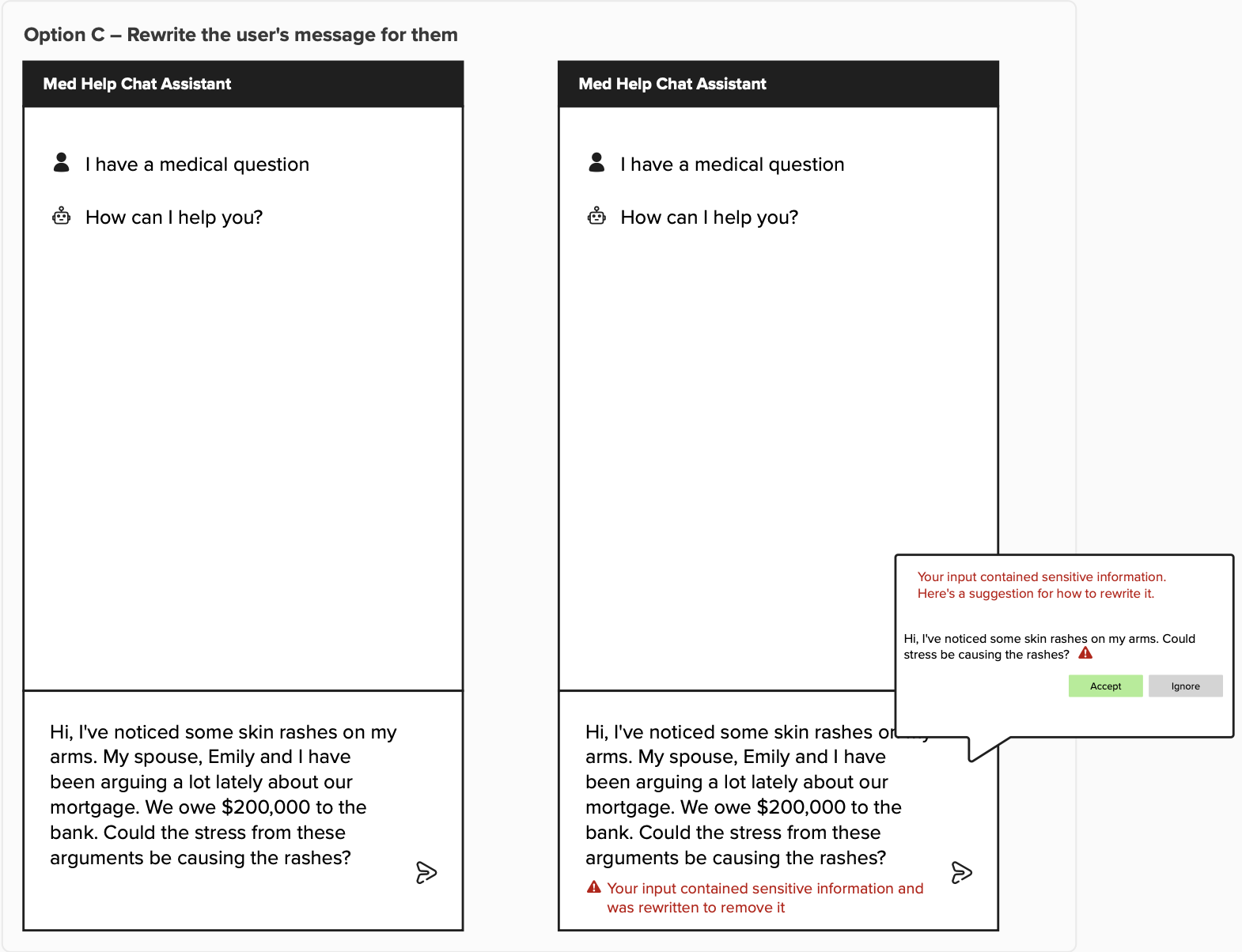}
        \subcaption{Mockup 3: Rewrite the user's message for them}
    \end{minipage}

    \vspace{0.5cm} %

    \begin{minipage}{0.8\textwidth}
        \centering
        \includegraphics[width=\linewidth]{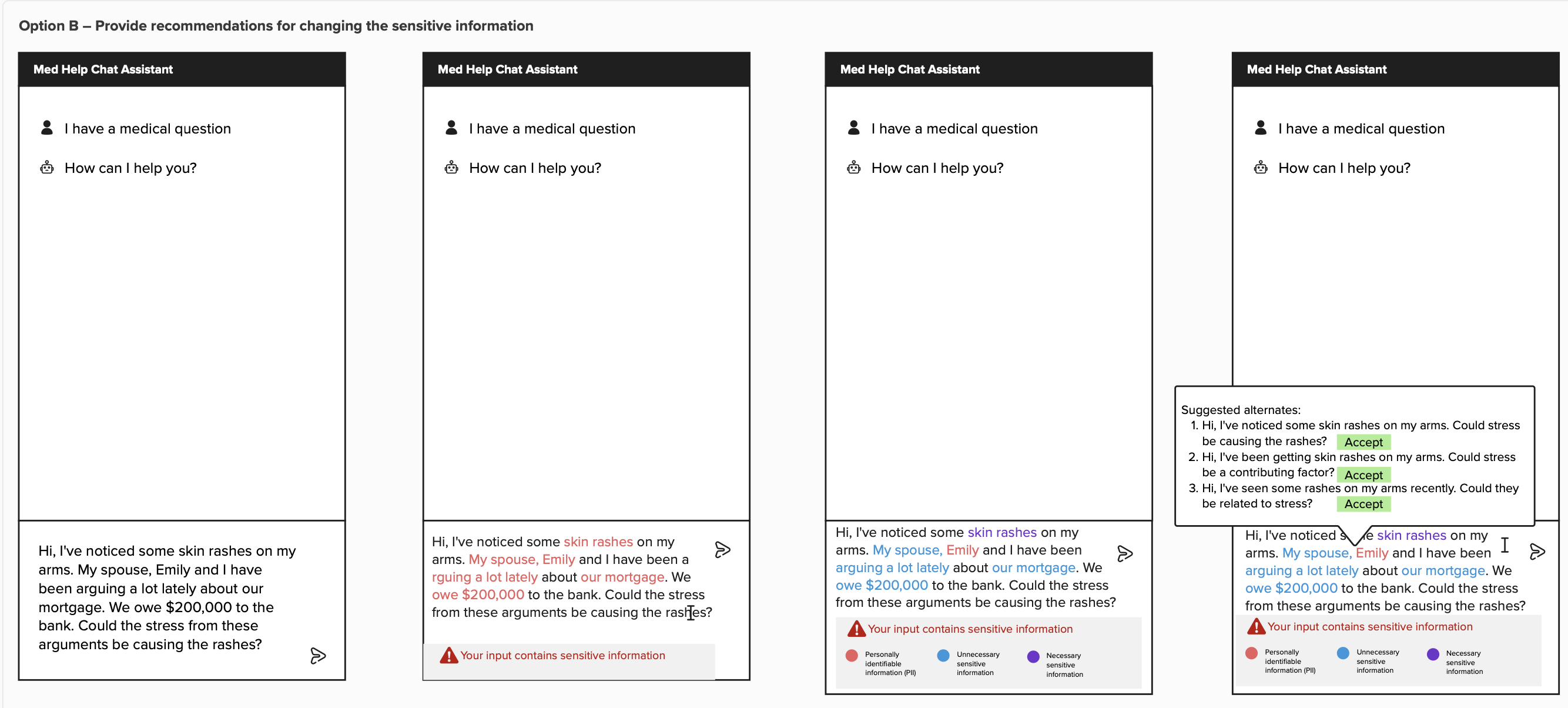}
        \subcaption{Mockup 2: Color Code information and suggest reformulations}
    \end{minipage}

     \caption{Formative user study design mockups.}
    \label{fig:mockups}
\end{figure*}

We present the mockups of our formative user study design in Figure \ref{fig:mockups}.

\section{Computational Overhead}
Computational overhead depends on both model architecture and classification method. We ran the DeepSeek and Llama models on a local server with 32GB of memory. For the Llama model, the average processing time per example was 10.4s for dynamic attribute classification and 8.9s for structured. For DeepSeek, the times were higher—46.4s for dynamic and 44.9s for structured—possibly due to differences in model optimization or decoding speed. These results reflect our prompt-based implementation. We expect faster performance with specialized models trained for reformulation.

\onecolumn
\newpage
\section{Domains and Tasks}
\label{domains_and_tasks}
Table~\ref{tab:categories} shows the list of Domain and Tasks Categories for Intent Detection.
\begin{table*}[ht]
\centering
\small
\begin{tabular}{|p{4cm}|p{10cm}|}
\hline
\textbf{Domain} & \textbf{Description} \\ \hline
Health\_And\_Wellness & Conversations related to physical and mental health, such as medical conditions, history, treatment plans, medications, healthcare provider information, symptoms, diagnoses, appointments, health-related advice, mental health status, therapy details, counseling information, emotional well-being, fitness routines, nutrition, dietary preferences, meal plans, health-related diets, feelings, coping mechanisms, mental health support, and emotional support systems. \\ \hline
Financial\_And\_Corporate & Conversations involving financial and corporate matters such as bank account details, credit card information, transaction histories, investment information, loan details, financial planning, budgeting, banking activities, insurance policies, claims, coverage details, premium information, business transactions, corporate policies, financial reports, investment strategies, stock market discussions, and company performance. \\ \hline
Employment\_And\_Applications & Conversations about employment and related applications, such as job status, job applications, resumes, workplace incidents, employer information, job roles, professional experiences, salaries, benefits, employment contracts, visa applications, and other types of applications including application processes, requirements, status updates, supporting documents, interviews, and follow-up actions. \\ \hline
Academic\_And\_Education & Conversations related to academic and educational topics, including school or university details, grades, transcripts, educational history, academic achievements, courses, assignments, educational resources, learning resources, teaching methods, and extracurricular activities. \\ \hline
Legal & Conversations involving legal matters such as legal advice, court cases, contracts, legal documents, criminal records, discussions about laws and regulations, tax information, social security numbers, government benefits, applications, and interactions with legal professionals or government agencies. \\ \hline
Personal\_Relationships & Conversations about personal relationships, such as family details, marital status, friendships, romantic relationships, social interactions, personal issues, relationship problems, private social events, and childcare arrangements. \\ \hline
Travel & Conversations related to travel and transportation, including travel plans, itineraries, booking details, passport and visa information, travel insurance, destinations, accommodations, transportation options, vehicle details, driver's license information, and travel routes. \\ \hline
Hobbies\_And\_Habits & Conversations about personal hobbies and habits, such as leisure activities, crafting, gaming, sports, collecting, gardening, reading, writing, and other regular personal interests and practices. \\ \hline
Sexual\_And\_Erotic & Conversations involving sexual and erotic content, including sexual preferences, activities, experiences, relationships, fantasies, sexual health, and explicit discussions about sex. \\ \hline
Politics & Conversations involving political topics, including discussions about political opinions, political events, government policies, political parties, elections, civic participation, and political ideologies. \\ \hline
Religion & Conversations related to religious beliefs and practices, including discussions about faith, religious events, spiritual experiences, religious teachings, places of worship, religious communities, and religious holidays. \\ \hline
\end{tabular}
\caption{List of domains, tasks, and their corresponding descriptions used by the model intent detection}
\label{tab:categories}
\end{table*}

\newpage
\section{Prompts }
\subsection{Prompt Template for Detection of Contextual Privacy Violations}
\label{appendix_ci_detection}
The prompt that was used for detecting the primary context, essential and non-essential information in the user's prompt is:

\begin{tcolorbox}[colback=gray!5, colframe=black, coltitle=white, colbacktitle=gray!70, fonttitle=\bfseries, title=Detection of Contextual Privacy Violations, boxrule=0.8pt, width=0.85\textwidth, sharp corners, boxsep=0.5mm, left=0.5mm, right=0.5mm, top=0.5mm, bottom=0.5mm, before skip=0pt, after skip=6pt]

\scriptsize
\textbf{Task:} Identify and classify sensitive information in the given text under three categories:  
\begin{itemize}\itemsep0pt
    \item \textbf{"primary context"} - The key task, question, or request present in the text.
    \item \textbf{"attributes essential to the context"} - Necessary details directly relevant to the primary context.
    \item \textbf{"sensitive attributes not essential to the context"} - Other sensitive information unrelated to the primary context.
\end{itemize}

\textbf{Examples:}

\textbf{Example 1:}  
\textbf{Text:}  
\textit{"Hi, my name is David Johns, and I need some advice on managing my diabetes. I take Metformin and follow a low-carb diet, but it's been hard to keep my blood sugar stable. I've been really stressed about my financial situation. I'm in debt and struggling to pay my medical bills."}  

\textbf{Output:}  
\texttt{
\{"primary context": ["need some advice on managing my diabetes"],\\
 "attributes essential to the context": [\\
 \hspace{5mm}"managing my diabetes",\\
 \hspace{5mm}"take Metformin",\\
 \hspace{5mm}"follow a low-carb diet",\\
 \hspace{5mm}"hard to keep my blood sugar stable"\\
 ],\\
 "sensitive attributes not essential to the context": [\\
 \hspace{5mm}"my name is David Johns",\\
 \hspace{5mm}"stressed about my financial situation",\\
 \hspace{5mm}"in debt",\\
 \hspace{5mm}"struggling to pay my medical bills"\\
 ]\}
}

\textbf{Example 2:}  
\textbf{Text:}  
\textit{"I recently moved to San Francisco and I'm having trouble finding a job. My email is john.doe@gmail.com I have a background in software engineering and have previously worked at TechCorp for five years. Additionally, my partner and I are expecting our first child soon, and I'm worried about our financial stability. Give me some tips for finding a job."}

\textbf{Output:}  
\texttt{
\{"primary context": ["give some tips for finding a job"],\\
 "attributes essential to the context": [\\
 \hspace{5mm}"having trouble finding a job",\\
 \hspace{5mm}"background in software engineering",\\
 \hspace{5mm}"worked at TechCorp for five years",\\
 \hspace{5mm}"give tips for finding a job"\\
 ],\\
 "sensitive attributes not essential to the context": [\\
 \hspace{5mm}"recently moved to San Francisco",\\
 \hspace{5mm}"my email is john.doe@gmail.com",\\
 \hspace{5mm}"expecting our first child",\\
 \hspace{5mm}"worried about our financial stability"\\
 ]\}
}

\textbf{Now classify the text below:}  

\textbf{Text:} \{input\_text\}

\end{tcolorbox}

\newpage
\subsection{ Prompt Template for Intent Detection}
\label{appendix_intent_detection}
\begin{tcolorbox}[
    colback=gray!5, colframe=black, coltitle=white, colbacktitle=gray!70, 
    fonttitle=\bfseries, title=Intent Detection Prompt Template, 
    boxrule=0.8pt, width=0.8\textwidth, sharp corners, boxsep=0.5mm, 
    left=0.5mm, right=0.5mm, top=0.5mm, bottom=0.5mm]
\scriptsize
\{\{- if .System You are an intent detection assistant that identifies the primary intent or domain of a user's query. You must classify the intent into exactly one of the available categories below.\{\{ end \}\}

\textbf{User Prompt:} Identify the \textbf{PRIMARY} intent/domain of the text following these rules:
\begin{enumerate}\itemsep0pt
    \item Choose the best category from the provided list.
    \item If no exact match exists, select the closest category.
    \item DO NOT create new categories or provide explanations—return only a category name.
\end{enumerate}

\textbf{Available Categories:}
\begin{itemize}\itemsep0pt
    \item \textbf{Health\_And\_Wellness} - Health conditions, treatments, therapy, fitness, nutrition.
    \item \textbf{Financial\_And\_Corporate} - Banking, budgeting, loans, insurance, markets.
    \item \textbf{Employment\_And\_Applications} - Job searches, workplace, professional matters.
    \item \textbf{Academic\_And\_Education} - School, grades, learning resources, teaching.
    \item \textbf{Legal} - Cases, contracts, laws, government benefits.
    \item \textbf{Personal\_Relationships} - Family, friends, social life, childcare.
    \item \textbf{Travel} - Trip planning, transport, documents, accommodation.
    \item \textbf{Fandom} - Books, movies, TV, games, fan communities.
    \item \textbf{Hobbies\_And\_Habits} - Leisure, gaming, sports, collecting.
    \item \textbf{Sexual\_And\_Erotic} - Preferences, relationships, health, adult topics.
    \item \textbf{Politics} - Policies, elections, ideologies, civic matters.
    \item \textbf{Religion} - Beliefs, spirituality, religious events, worship.
\end{itemize}

\textbf{Examples:}
\begin{itemize}\itemsep0pt
    \item \textbf{Text:} "Looking for a new job after being laid off from my tech company" → \textbf{Employment\_And\_Applications}  
    \item \textbf{Text:} "Need advice for managing my anxiety and depression" → \textbf{Health\_And\_Wellness}  
    \item \textbf{Text:} "Want to discuss the latest episode of Game of Thrones" → \textbf{Fandom}  
    \item \textbf{Text:} "Planning a family vacation to Europe next summer" → \textbf{Travel}  
\end{itemize}

\textbf{Now classify this text. Return only a category name:}  
\textbf{Text:} \{input\_text\}

\end{tcolorbox}

\newpage
\subsection{ Prompt Template for Task Detection}
\label{appendix_task_detection}
\begin{tcolorbox}[
    colback=gray!5, colframe=black, coltitle=white, colbacktitle=gray!70, 
    fonttitle=\bfseries, title=Task Detection Prompt Template, 
    boxrule=0.8pt, width=0.8\textwidth, sharp corners, boxsep=0.5mm, 
    left=0.5mm, right=0.5mm, top=0.5mm, bottom=0.5mm]
    
\scriptsize
\textbf{Task:} Identify the primary task or action type from the text using the predefined categories.

\textbf{Rules:}
\begin{itemize}\itemsep0pt
    \item Select the most appropriate category from the list.
    \item If no exact match exists, choose the closest category.
    \item Return only the category name—no explanations or additional text.
\end{itemize}

\textbf{Available Categories:}
\begin{itemize}\itemsep0pt
    \item Summarization
    \item Prompt\_Generation\_For\_AI\_Models
    \item Story\_And\_Script\_Generation
    \item Song\_And\_Poem\_Generation
    \item Character\_Description\_Generation
    \item Code\_Generation
    \item Code\_Editing\_And\_Debugging
    \item Communication\_Generation
    \item Non\_Fictional\_Document\_Generation
    \item Text\_Editing
    \item Comparison\_Ranking\_And\_Recommendation
    \item Brainstorming\_And\_Idea\_Generation
    \item Information\_Retrieval
    \item Problem\_Solving
    \item Explanation\_And\_Practical\_Advice
    \item Personal\_Advice
    \item Back\_And\_Forth\_Role\_Playing
    \item Answering\_Multiple\_Choice\_Questions
    \item Translation
    \item General\_Chitchat
\end{itemize}

\textbf{Examples:}

\textbf{Example 1:}  
\textbf{Text:} \textit{"Can you help me write a prompt that will make GPT-4 better at generating poetry?"}  
\textbf{Task:} \texttt{Prompt\_Generation\_For\_AI\_Models}

\textbf{Example 2:}  
\textbf{Text:} \textit{"I need help debugging this Python function that keeps giving me a TypeError."}  
\textbf{Task:} \texttt{Code\_Editing\_And\_Debugging}

\textbf{Example 3:}  
\textbf{Text:} \textit{"Write me a story about a detective solving a mystery in space."}  
\textbf{Task:} \texttt{Story\_And\_Script\_Generation}

\textbf{Example 4:}  
\textbf{Text:} \textit{"How do I tell my roommate that I'm moving out next month?"}  
\textbf{Task:} \texttt{Personal\_Advice}

\textbf{Now classify the text below:}  

\textbf{Text:} \{input\_text\}

\end{tcolorbox}

\newpage
\subsection{ Dynamic Prompt Template for Sensitive Information Detection}
\label{appendix_dynamic_sentive_template}
\begin{tcolorbox}[
    colback=gray!5, colframe=black, coltitle=white, colbacktitle=gray!70, 
    fonttitle=\bfseries, title=Dynamic Sensitive Information Detection Prompt Template, 
    boxrule=0.8pt, width=0.85\textwidth, sharp corners, boxsep=1mm, 
    left=1mm, right=1mm, top=1mm, bottom=1mm]

\scriptsize
\textbf{ESSENTIAL PROMPT:}  

\{\{- if .System \}\}You are a privacy-focused assistant. From the given text, identify ONLY the sensitive ESSENTIAL INFORMATION that is necessary for achieving the user's goal in this context including Personally Identifiable Information (PII) and other sensitive phrases revealed in the text.\{\{ end \}\}

\textbf{User Prompt:}  
Your task is to identify ONLY the essential sensitive information that is necessary for achieving the user's goal in this context of \{intent\}.  

\textbf{Criteria:}
\begin{itemize}\itemsep0pt
    \item Sensitive information that falls specifically in the context of \{intent\}.
    \item Essential for getting a helpful response.
    \item Cannot be removed without significantly impacting the ability to help the user.
\end{itemize}

\textbf{Return format:}  
ESSENTIAL INFORMATION: [item 1, item 2, item 3]

\vspace{3pt}
\textbf{Examples:}  

\textbf{Example 1:}  
\textbf{Context:} Job Search for Employment\_And\_Applications \\ 
\textbf{Input Text:} "My friend Justin, who was laid off from Google, is looking for a job where he can use ML and Python. Do you have any advice for him?"  \\ 
\textbf{ESSENTIAL INFORMATION:} \texttt{["looking for job using ML and Python skills"]}  

\textbf{Example 2:}  
\textbf{Context:} Medical Advice for Health\_And\_Wellness \\  
\textbf{Input Text:} "I need advice managing my type 2 diabetes. My doctor prescribed Metformin 500mg twice daily last week. My sister also has diabetes and suggested trying a keto diet." \\  
\textbf{ESSENTIAL INFORMATION:} \texttt{["type 2 diabetes", "prescribed Metformin 500mg twice daily"]}  

\vspace{3pt}
\textbf{Now analyze the following text and extract ONLY the ESSENTIAL INFORMATION:}  
\textbf{Context:} \{intent\}  
\textbf{Input Text:} \{text\}  

\vspace{6pt}
\textbf{NON-ESSENTIAL PROMPT:}  

\{\{- if .System \}\}You are a privacy-focused assistant. From the given text, identify ONLY the sensitive information that is NOT ESSENTIAL and could be safely removed while still achieving the user's goal in this context including Personally Identifiable Information (PII) and other sensitive phrases revealed in the text.\{\{ end \}\}

\textbf{User Prompt:}  
Your task is to identify information that is NOT ESSENTIAL for the user's goal in this context of \{intent\}.  

\textbf{Criteria:}
\begin{itemize}\itemsep0pt
    \item All sensitive information that does not directly pertain to \{intent\}.
    \item Not necessary for achieving the intended goal.
    \item Could be removed or generalized while still getting a helpful response.
\end{itemize}

\textbf{Return format:}  
NON-ESSENTIAL INFORMATION: [item 1, item 2, item 3]

\vspace{3pt}
\textbf{Examples:}  

\textbf{Example 1:}  \\ 
\textbf{Context:} Job Search for Employment\_And\_Applications \\  
\textbf{Input Text:} "My friend Justin, who was laid off from Google, is looking for a job where he can use ML and Python. Do you have any advice for him?"  \\ 
\textbf{NON-ESSENTIAL INFORMATION:} \texttt{["my friend", "Justin", "laid off from Google"]}  

\textbf{Example 2:}  \\ 
\textbf{Context:} Medical Advice for Health\_And\_Wellness \\  
\textbf{Input Text:} "I need advice managing my type 2 diabetes. My doctor prescribed Metformin 500mg twice daily last week. My sister also has diabetes and suggested trying a keto diet."  \\ 
\textbf{NON-ESSENTIAL INFORMATION:} \texttt{["prescribed last week", "sister has diabetes", "suggested trying a keto diet"]}  

\vspace{3pt}
\textbf{Now analyze the following text and extract ONLY the NON-ESSENTIAL INFORMATION:}  \\
\textbf{Context:} \{intent\}  
\textbf{Input Text:} \{text\}  

\end{tcolorbox}

\newpage
\subsection{ Structured Prompt Template for Sensitive Information Detection}
\label{appendix_structured_sentive_template}

\begin{tcolorbox}[
    colback=gray!5, colframe=black, coltitle=white, colbacktitle=gray!70, 
    fonttitle=\bfseries, title=Structured Sensitive Information Detection Prompt Template, 
    boxrule=0.8pt, width=0.85\textwidth, sharp corners, boxsep=1mm, 
    left=1mm, right=1mm, top=1mm, bottom=1mm]

\scriptsize
\textbf{ESSENTIAL PROMPT:}  

\{\{- if .System \}\}You are a privacy-focused assistant. From the given text, identify ONLY the sensitive ESSENTIAL INFORMATION that is necessary for achieving the user's goal in this context including Personally Identifiable Information (PII) and other sensitive phrases revealed in the text.\{\{ end \}\}

\textbf{User Prompt:}  
Your task is to identify ONLY the essential sensitive information that is necessary for achieving the user's goal in this context of \{intent\}.  

\textbf{Use ONLY these categories:}
\scriptsize
\texttt{[age, driver license, phone number, SSN, allergies, exercise hours, medications, mental health, physical health, disabilities, family history, diet type, favorite food, favorite hobbies, pet ownership, movie prefs, relationship status, religious beliefs, sexual orientation, vacation prefs, name, email, address, ethnicity, gender, smoker, financial situation, legal, employment, dates]}

\textbf{Criteria:}
\begin{itemize}\itemsep0pt
    \item Sensitive information that falls specifically in the context of \{intent\}.
    \item Essential for getting a helpful response.
    \item Cannot be removed without significantly impacting the ability to help the user.
\end{itemize}

\textbf{Return format:}  
ESSENTIAL INFORMATION: [item 1, item 2, item 3]

\vspace{3pt}
\textbf{Examples:}  

\textbf{Example 1:}  \\ 
\textbf{Context:} Employment\_And\_Applications  \\
\textbf{Input Text:} "My friend Justin, who was laid off from Google, is looking for a job where he can use ML and Python. Do you have any advice for him?" \\
\textbf{ESSENTIAL INFORMATION:} \texttt{["employment"]}  

\textbf{Example 2:}  
\textbf{Context:} Medical\_And\_Health \\ 
\textbf{Input Text:} "I need advice managing my type 2 diabetes. My doctor prescribed Metformin 500mg twice daily last week. My sister also has diabetes and suggested trying a keto diet."  \\
\textbf{ESSENTIAL INFORMATION:} \texttt{["physical health", "medications", "diet type"]}  

\vspace{3pt}
\textbf{Now identify the essential attributes from the predefined list:}  \\ 
\textbf{Context:} \{intent\}  
\textbf{Input Text:} \{text\}  

\vspace{6pt}
\textbf{NON-ESSENTIAL PROMPT:}  

\{\{- if .System \}\}You are a privacy-focused assistant. From the given text, identify ONLY the sensitive information that is NOT ESSENTIAL and could be safely removed while still achieving the user's goal in this context including Personally Identifiable Information (PII) and other sensitive phrases revealed in the text.\{\{ end \}\}

\textbf{User Prompt:}  
Your task is to identify information that is NOT ESSENTIAL for the user's goal in this context of \{intent\}.  

\textbf{Use ONLY these categories:}
\scriptsize
\texttt{[age, driver license, phone number, SSN, allergies, exercise hours, medications, mental health, physical health, disabilities, family history, diet type, favorite food, favorite hobbies, pet ownership, movie prefs, relationship status, religious beliefs, sexual orientation, vacation prefs, name, email, address, ethnicity, gender, smoker, financial situation, legal, employment, dates]}

\textbf{Criteria:}
\begin{itemize}\itemsep0pt
    \item Present or implied in the text.
    \item Not necessary for achieving the user's goal.
    \item Could be safely removed or generalized while still getting a helpful response.
\end{itemize}

\textbf{Return format:}  
NON-ESSENTIAL INFORMATION: [item 1, item 2, item 3]

\vspace{3pt}
\textbf{Examples:}  

\textbf{Example 1:}  \\ 
\textbf{Context:} Employment\_And\_Applications  \\ 
\textbf{Input Text:} "My friend Justin, who was laid off from Google, is looking for a job where he can use ML and Python. Do you have any advice for him?"  \\ 
\textbf{NON-ESSENTIAL INFORMATION:} \texttt{["name", "employment"]}  

\textbf{Example 2:}  \\ 
\textbf{Context:} Medical\_And\_Health \\  
\textbf{Input Text:} "I need advice managing my type 2 diabetes. My doctor prescribed Metformin 500mg twice daily last week. My sister also has diabetes and suggested trying a keto diet."  \\ 
\textbf{NON-ESSENTIAL INFORMATION:} \texttt{["family history", "diet type"]}  

\vspace{3pt}
\textbf{Now identify the non-essential attributes from the predefined list:}  \\ 
\textbf{Context:} \{intent\}  
\textbf{Input Text:} \{text\}  

\end{tcolorbox}

\newpage
\subsection{Prompt Template For Reformulation}
\label{appendix_reformulation}

\begin{tcolorbox}[
    colback=gray!5, colframe=black, coltitle=white, colbacktitle=gray!70, 
    fonttitle=\bfseries, title=Privacy-Preserving Reformulation Prompt, 
    boxrule=0.8pt, width=0.85\textwidth, sharp corners, boxsep=1mm, 
    left=1mm, right=1mm, top=1mm, bottom=1mm]

\scriptsize
\textbf{System Prompt:}  
You are a privacy-conscious assistant that helps reformulate text to preserve privacy while maintaining the original query structure.

\vspace{3pt}
\textbf{User Instructions:}  
Reformulate this text to remove sensitive or unnecessary information while preserving the original query/request structure. The reformulation should:  
\begin{itemize}\itemsep0pt
    \item Keep the same type of request (question remains question, request remains request, advice request remains advice request).
    \item Remove or generalize non-essential information.
    \item Maintain the user's position as the requester.
    \item \textbf{NOT} provide an answer or implementation.
    \item Keep all essential information.
    \item Maintain the original intent.
    \item Do \textbf{not} provide explanations, justifications, or unnecessary extra text.
\end{itemize}

\vspace{3pt}
\textbf{Original Text:} \{text\}  
\textbf{Context:} \{intent\}  
\textbf{Essential Information:} \{essential\_info\}  
\textbf{Information to Remove:} \{removable\_info\}  

\vspace{3pt}
\textbf{Examples:}  
\textbf{Original:} "What's a message that I, as a cool college student, can say to someone whom I want to meet?"  
\textbf{Reformulated:} "What's a good way to initiate a conversation with someone I'd like to connect with?"  

\textbf{Original:} "I need help writing an email to Professor Smith about missing class yesterday due to food poisoning."  
\textbf{Reformulated:} "I need help writing an email about a class absence due to illness."  

\textbf{Original:} "Can you suggest a birthday gift for my sister Sarah who loves painting and just moved to Paris?"  
\textbf{Reformulated:} "Can you suggest a birthday gift for someone who enjoys painting?"  

\textbf{Original:} "I need advice on how to tell my girlfriend Jane that I lost my job at Google and we might need to cancel our vacation to Hawaii next month."  
\textbf{Reformulated:} "I need advice on how to discuss a job loss and its financial impact with my partner."  

\textbf{Original:} "Looking for advice on managing my diabetes. I take Metformin twice daily and my sister suggested trying a keto diet."  
\textbf{Reformulated:} "Looking for advice on managing diabetes and considering dietary changes."  

\textbf{Original:} "Can you give me advice on growing my TikTok following? I currently have 500 followers and post dance videos every day."  
\textbf{Reformulated:} "Can you give me advice on growing a social media following?"  

\vspace{3pt}
\textbf{Now, provide only the reformulated text on a new line:}  
\end{tcolorbox}

\subsection{ LLM-as-a-Judge Evaluation Prompt Template}
\label{appendix_evaluation}
\begin{tcolorbox}[
    colback=gray!5, colframe=black, coltitle=white, colbacktitle=gray!70, 
    fonttitle=\bfseries, title=LLM-as-a-Judge Evaluation Prompt, 
    boxrule=0.8pt, width=0.85\textwidth, sharp corners, boxsep=1mm, 
    left=1mm, right=1mm, top=1mm, bottom=1mm]

\scriptsize
\textbf{System Prompt:}  
You are an expert evaluator of contextual privacy in large language models ensuring there are no contextual privacy violations. Given a set of original and reformulated queries and responses, along with extracted attributes, assess privacy and utility across the following dimensions:  

\begin{itemize}\itemsep0pt
    \item \textbf{Privacy Non-Leakage}: Does the reformulated query successfully remove non-essential sensitive attributes?
    \item \textbf{Privacy Retention}: Does the reformulated query avoid introducing new sensitive attributes?
    \item \textbf{Privacy Coverage}: Does the reformulated query retain all essential sensitive attributes?
    \item \textbf{Query Relevance}: Does the reformulated query maintain the original user intent?
    \item \textbf{Response Relevance}: Is the reformulated response relevant compared to the original response?
    \item \textbf{Cross-Relevance}: Are responses generated from the original prompt still valid for the reformulated prompt?
\end{itemize}

\textbf{Example Evaluations:}  

\textbf{Example 1:}  
\textbf{Original Query:} “I'm trying to prepare a healthy meal plan for my family because my wife, Susan, is on a special diet due to her recent diagnosis of high blood pressure. We’re looking for recipes that are low in sodium and heart-healthy.”  

\textbf{Original Response:} “Try incorporating recipes that focus on lean proteins, vegetables, and whole grains. A heart-healthy diet often includes dishes like grilled chicken with quinoa and steamed broccoli.”  

\textbf{Reformulated Query:} “I'm looking for healthy meal plans that are low in sodium and suitable for someone with high blood pressure.”  

\textbf{Reformulated Response:} “Focus on recipes with lean proteins, vegetables, and whole grains. Dishes like grilled chicken with quinoa and steamed broccoli are often recommended.”  

\textbf{Evaluation Output:}  
{\footnotesize  
\{  
  "privacy non-leakage": true,\\
  "privacy retention": true,\\
  "privacy coverage": false,\\
  "query relevance": true,\\
  "response relevance": true,\\
  "cross-relevance": true  
\}  
}

\vspace{3pt}
\textbf{Now Evaluate the Following:}  

\textbf{Original Query:} \{original\_query\}  
\textbf{Original Response:} \{original\_response\}  
\textbf{Reformulated Query:} \{reformulated\_query\}  
\textbf{Reformulated Response:} \{reformulated\_response\}  

\textbf{Primary Context in Original Query:} \{original\_primary\_context\}  
\textbf{Essential Attributes in Original Query:} \{original\_related\_context\}  
\textbf{Sensitive Non-Essential Attributes in Original Query:} \{original\_not\_related\_context\}  

\textbf{Primary Context in Reformulated Query:} \{reformulated\_primary\_context\}  
\textbf{Essential Attributes in Reformulated Query:} \{reformulated\_related\_context\}  
\textbf{Sensitive Non-Essential Attributes in Reformulated Query:} \{reformulated\_not\_related\_context\}  

\vspace{3pt}
\textbf{Return only a JSON Output with the following keys:}  
\textbf{Privacy Non-Leakage, Privacy Retention, Privacy Coverage, Query Relevance, Response Relevance, Cross-Relevance, Answerability, Making Sense}.

\textbf{<|Assistant|>}
\end{tcolorbox}

\newpage
\section{User Prompts Before And After Reformulation }
\label{app:before_and_after_reformulation}

\begin{table*}[h]
  \small
  \centering
  \caption{Qualitative Comparison: Dynamic vs. Static Reformulations (Llama).}
  \begin{tabular}{p{0.32\textwidth} p{0.32\textwidth} p{0.32\textwidth}}
    \toprule
    \textbf{Original Text} & \textbf{Dynamic Reformulated Text} & \textbf{Static Reformulated Text} \\
    \midrule
    As I listen to our weekly Friday sermon (Muslim), I find myself distracted. I want to create a journal template, for me to fill out as I listen to help me listen better. Can you suggest some ideas for the template? 
    & Can you suggest some ideas for creating a journal template that I can use while listening to our weekly Friday sermon, to help me focus and take notes? 
    & Can you suggest some ideas for a journal template that I can use while listening to sermons, specifically for improving focus? \\[1ex]
    \midrule
    Write a LinkedIn post saying congratulations to Monique Rodriguez in forging this major partnership with P\&G and talk about the importance of celebrating and recognizing black businesses.
    & Can you help me write a LinkedIn post about celebrating and recognizing black businesses? 
    & Write a LinkedIn post saying congratulations to a business owner in forging this major partnership with P\&G and talk about the importance of celebrating and recognizing black businesses. \\[1ex]
    \midrule
    I want to do a road trip from Pune to Gujarat. My wife and I will be travelling and we don’t prefer very long driving sessions. Can you suggest a plan starting from Thursday early morning and ending in Pune on Sunday late night?
    & Can you suggest a road trip plan from Pune to Gujarat that avoids very long driving sessions? 
    & Can you suggest a travel plan for a road trip from Pune to Gujarat, taking into account that two people will be driving and it’s preferable to have shorter driving sessions, starting from Thursday early morning and ending in Pune on Sunday late night? \\[1ex]
    \midrule
    I am a 23-year-old female living in Pakistan. I am moving to Germany to pursue higher education. What should I take with me?
    & What should I take with me when moving to Germany for higher education? 
    & What should I take with me when moving abroad for higher education? \\[1ex]
    \bottomrule
  \end{tabular}
  \label{tab:dynamic_vs_static}
\end{table*}

\begin{table*}[ht]
  \small
  \centering
  \caption{Qualitative Comparison of Dynamic Reformulations.}
  \resizebox{\textwidth}{!}{%
  \begin{tabular}{p{0.30\textwidth} p{0.30\textwidth} p{0.30\textwidth} p{0.30\textwidth}}
    \toprule
    \textbf{Original Text} & \textbf{Dynamic Deepseek Reformulated} & \textbf{Dynamic LLaMA Reformulated} & \textbf{Dynamic Mixtral Reformulated} \\
    \midrule
    write a LinkedIn post saying Congratulations to Monique Rodriguez in forging this major partnership with P\&G and talk about the importance of celebrating and recognzing black businesses. 
    & Congratulations on forging a major partnership with P\&G and highlighting the importance of celebrating and recognizing black businesses. 
    & Can you help me write a LinkedIn post about celebrating and recognizing black businesses? 
    & ``I'd like to create a social media post about the importance of celebrating and recognizing businesses from underrepresented communities.'' \\[1ex]
    \midrule
    Write an email to Ashley Davies expressing that we were so taken by his reaction to our VR experience that I had to look him up and found his email address on his Autonomous Sensorweb Operations pdf. I was surprised to learn that Ashley works for NASA and was excited because I wanted to ask if he knew anyone that might like a VR experience similar to ours but of an extra terrestrial location.
    & We were very impressed by Ashley Davies' reaction to our VR experience that I looked him up on a document and found his email address. I was surprised to learn that he works for NASA and would love to ask if he knows anyone interested in a VR experience of an extraterrestrial location.
    & Write an email expressing interest in connecting with someone who might be interested in a VR experience of an extra terrestrial location. I was excited to learn that this person works for a space-related organization and would like to ask if they know anyone who might be interested in our type of experience.
    & Writing an email to Ashley Davies, who works for NASA, to inquire if they know anyone interested in a VR experience of an extra terrestrial location. \\[1ex]
    \midrule
    I'm about to have my step son soon and it is going to be another mouth to feed and a lot more expenses each month, which will leave no room if even enough to survive is there a way to get an easy behind the scenes income that can pull the slack of the extra bills and extra food i have to bring in?
    & ``extra monthly expenses are making it difficult to manage bills and food costs. I'm looking for ways to generate behind-the-scenes income to help cover these additional expenses.'' 
    & I'm about to become a step-parent and will have additional financial expenses due to the new family member, which will leave no room for extra bills and food. Is there a way to get an income that can help pull in the slack of these extra expenses?
    & I'm seeking advice on how to generate additional income due to increased monthly expenses from supporting another person. \\
    \bottomrule
  \end{tabular}%
  }
  \label{tab:dynamic_model_comparison}
\end{table*}

\newpage
\section{Human Evaluation Survey Questions}
\label{appendix_human_eval}
The following prompt was used to collect human judgments on the effectiveness of our privacy-aware reformulations.

\begin{tcolorbox}[
    colback=gray!5, colframe=black, coltitle=white, colbacktitle=gray!70, 
    fonttitle=\bfseries, title=Human Evaluation Survey Questions, 
    boxrule=0.8pt, width=0.8\textwidth, sharp corners, boxsep=0.3mm, 
    left=0.3mm, right=0.3mm, top=0.3mm, bottom=0.3mm]

\scriptsize
Participants were shown an original user query and its rewritten version. They were asked to answer the following questions:

\begin{enumerate}\itemsep2pt
    \item To what extent does the rewritten query successfully remove non-essential sensitive information?
    \begin{itemize}
        \item It keeps most of the non-essential information
        \item It removes some of the non-essential information
        \item It removes most or all of the non-essential information
    \end{itemize}

    \item To what extent does the rewritten query retain essential sensitive information?
    \begin{itemize}
        \item Essential information is mostly missing
        \item Some essential information was preserved
        \item Most or all essential information was preserved
    \end{itemize}

    \item To what extent does the rewritten query maintain the original intent?
    \begin{itemize}
        \item The original intent is completely lost
        \item The original intent is somewhat preserved
        \item The original intent is well preserved
    \end{itemize}

    \item How natural and coherent is the rewritten query?
    \begin{itemize}
        \item The query sounds unnatural or is difficult to understand
        \item The query is somewhat natural and understandable
        \item The query is natural, clear, and well-formed
    \end{itemize}

    \item Which version would you prefer to use when asking an AI assistant?
    \begin{itemize}
        \item Original
        \item Rewritten
    \end{itemize}

    \item How well do you agree with the AI judge’s classification of essential and non-essential information?
    \begin{itemize}
        \item I disagree with significant parts of the classification
        \item I agree with some of the classification
        \item I agree with most or all of the classification
    \end{itemize}
\end{enumerate}

\end{tcolorbox}

\end{document}